\documentclass{aa} 

\usepackage{graphicx}
\usepackage{txfonts}
\usepackage{appendix}
\usepackage{mathrsfs}

\begin{document}
\newcommand{\HD}{HD\,50138}
\newcommand{\HI}{\ion{H}{i}}
\newcommand{\brg}{Br$\gamma$}
\newcommand{\NaI}{\ion{Na}{i}}
\newcommand{\macc}{$\dot{M}_{acc}$}
\newcommand{\lacc}{L$_{acc}$}
\newcommand{\lbol}{L$_{bol}$}
\newcommand{\kms}{km\,s$^{-1}$}
\newcommand{\um}{$\mu$m}
\newcommand{\lam}{$\lambda$}
\newcommand{\msyr}{M$_{\odot}$\,yr$^{-1}$}
\newcommand{\Av}{A$_V$}
\newcommand{\msun}{M$_{\odot}$}
\newcommand{\Lsun}{L$_{\odot}$}
\newcommand{\Lstar}{L$_{\star}$}
\newcommand{\Tstar}{T$_{\star}$}
\newcommand{\egs}{erg\,s$^{-1}$\,cm$^{-2}$\,\AA$^{-1}$}
\newcommand{\simless}{\mathbin{\lower 3 pt\hbox {$\rlap{\raise 5pt\hbox{$\char'074$}}\mathchar"7218$}}} 
\newcommand{\simgreat}{\mathbin{\lower 3pt\hbox {$\rlap{\raise 5pt\hbox{$\char'076$}}\mathchar"7218$}}} 

   \title{The circumstellar environment of HD\,50138 revealed by VLTI/\textit{AMBER} at high angular resolution 
   \thanks{Based on observations collected at the European Southern Observatory Paranal, Chile (ESO programme 096.C-0293(A)).}
   }


    \author{M. Koutoulaki 
          \inst{1,2}
          \and
          R. Garcia Lopez\inst{1}
          \and
          A. Natta\inst{1,3}
             \and
          A. Caratti o Garatti\inst{1}
             \and
          D. Coffey\inst{2,1}
            \and
            J. Sanchez-Bermudez\inst{4}
            \and
          T. P. Ray\inst{1}
          }

  \institute{Dublin Institute for Advanced Studies, 31 Fitzwilliam Place, Dublin 2, Ireland\\
              \email{mariakout@cp.dias.ie}\\
         \and
             School of Physics, University College Dublin, Belfield, Dublin 4, Ireland\\
                \and
                     INAF/Osservatorio Astrofisico di Arcetri, Largo E. Fermi 5, 50125 Firenze, Italy\\
                     \and
                     European Southern Observatory, Alonso de Cordova 3107, Vitacura, Santiago, Chile \\
                    }

   \date{Received February 12, 2018; accepted April 23, 2018}

 
  \abstract
   {\HD\ is a Herbig B[e] star with a circumstellar disc detected at infrared and millimeter wavelength. Its brightness makes it a good candidate for near-infrared interferometry observations.} 
   { We aim to resolve, spatially and spectrally, the continuum and hydrogen emission lines in the 2.12-2.47 micron region, to shed light on the immediate circumstellar environment of the star.}
  { Very Large Telescope-Interferometer/\textit{AMBER} (VLTI/\textit{AMBER}) K-band observations provide spectra, visibilities, differential phases, and closure phases along three long baselines for the continuum, and \HI\ emission in \brg\ and five high-n Pfund lines. By computing the pure-line visibilities, we derive the angular size of the different line-emitting regions. A simple local thermodynamic equilibrium (LTE) model was created to constrain the physical conditions of \HI\ emitting region.}  
   {The continuum region cannot be reproduced by a geometrical two-dimensional (2D) elongated Gaussian fitting model. We estimate the size of the region to be 1 au. We find the detected hydrogen lines (\brg\ and Pfund lines) come from a more compact region of size 0.4 au. The \brg\ line exhibits an S-shaped differential phase, indicative of rotation. The continuum and \brg\ line closure phase show offsets of $\sim$-25$\pm$5 \degr\ and 20$\pm$10\degr\, respectively. This is evidence of an asymmetry in their origin, but with opposing directions. We find that we cannot converge on constraints for the \HI\ physical parameters without a more detailed model.} 
  {Our analysis reveals that \HD\ hosts a complex circumstellar environment. Its continuum emission cannot be reproduced by a simple disc brightness distribution. 
 Similarly, several components must be evoked to reproduce the interferometric observables within the \brg\, line. Combining the spectroscopic and interferometric data of the \brg\ and Pfund lines favours an origin in a wind region with a large opening angle. Finally, although we cannot exclude the possibility that \HD\ is a young star our results point to an evolved source.} 


\titlerunning{The circumstellar environment of \HD\, revealed using VLTI}
\authorrunning{Koutoulaki et al.}
   \maketitle
%

\section{Introduction}

The evolutionary nature of Herbig B[e] stars is as yet unclear. They are intermediate mass stars (2-10 M$_{\sun}$), showing permitted and forbidden emission lines as well as a large infrared excesses, but they may be either pre-main sequence or evolved stars. 
Although there are many studies addressing the properties of the dusty outer disc, there are only a few on the gas component of the disc, in particular of its inner structure. This requires  milli-arcsecond resolution, and so far there have been only a few cases where Herbig Ae/Be and evolved B[e] stars at high spatial and high spectral resolution have been studied \citep[e.g.][]{Kraus2008,Weigelt2011,Alessio2015,Rebeca2015,Rebeca2016}.
These studies have mostly focused on the \HI\ \brg\ line, which is the brightest emission line in the K band in Herbig stars. The emission of this line was initially associated with the accretion process through an empirical relation \citep{Muzerolle1998,Calvet2004} but recent near-infrared (NIR) interferometric observations have shown that it can in part be emitted in a wind \citep{Weigelt2011,Rebeca2015,Rebeca2016,Alessio2015}.

The star HD50138 (MWC158, V743 Mon, IRAS 06491-0654) is a Herbig B[e] star and is one of the brightest B[e] stars in the southern sky. Its stellar parameters are shown in Table \ref{tab:param}. It is  not associated with any star-forming region but it might be part of the Orion Monoceros molecular cloud complex \citep{Maddalena1986}. It is located at a distance of 340$\pm$47 pc \citep{GAIA2016}. The evolutionary status of this source is not clear. On the one hand,  the star has been considered as a pre-main sequence object \citep{Morrison1995}. On the other hand, it has also been classified as an evolved object close to turn-off from the main sequence \citep{Fernandes2009}. A reason to consider it a young star is because some infall signatures in the spectrum are present, but the fact that it is located far away from a star-forming region makes this criterion alone suspect. Moreover, its observed variability in the UV (ultraviolet) \citep{Hutsemekers1985}, which is associated with an outburst or a shell phase \citep{Andrillat1991}, is more often found in post-main sequence evolution.\\
\HD\ has a spectral energy distribution (SED) typical of Herbig Be stars \citep{Ellerbroek2015}. 
Polarimetry studies find evidence of a circumstellar rotating disc \citep{Oudmaijer1999,Harrington2007,Harrington2009}. The inner disc regions have been imaged using NIR interferometry and found to be highly asymmetric \citep{Lazareff2016,Kluska2016}, possibly changing with time \citep{Kluska2016}. \HD\ has a strong \brg\ emission, which has also been spatially resolved with interferometry, and it appears to originate in the inner disc region \citep[ with size of 1.5 mas,][]{Ellerbroek2015}.\\
In this paper, we present NIR interferometric data of the innermost circumstellar environment of \HD. The observations and data reduction are discussed in Section 2 and the results, discussions, and conclusion in Sections 3, 4 and 5 respectively.

\begin{table}[h]
\caption{Stellar properties of \HD\ }             
\label{tab:param}      
\centering                          
\begin{tabular}{l l c}        
\hline\hline                 
Parameters & Value & Reference\\    
\hline  
   RA (2000)& 06 51 33.40 &\\
   DEC(2000) & -06 57 59.5 &\\
   spectral type & B8  &1\\
   K (mag) & 4.32 mag  &2\\
   mass & 6 $M_{\odot}$  &3, 5\\      
   d & $340 \pm 47$ pc  &4\\  
   $L_{*}$ &$(1.2\pm 0.4) \times 10^3 L_{\odot}$  &5\\
  $R_{*}$ &$7.0\pm 2.1 R_{\odot}$ & 5\\
  T& 13000 K  & 5\\
   
\hline                                   
\end{tabular} 
\tablefoot{(1)\, \citet{Gray1998}, (2)\, \citet{Lazareff2016}, (3)\, \citet{Fernandes2009}, (4)\, \citet{GAIA2016}, (5)\, \citet{Ellerbroek2015} }
\end{table}


\section{Observations and data reduction}

HD\,50138 was observed using the VLT-Interferometer (VLTI) with the beam combiner \textit{AMBER} (Astronomical Multi-BEam combineR) \citep{Petrov2007} on the 2015 December 23. The observations were performed using the 8.2\,m unit telescopes (UT) with the configuration UT1-UT3-UT4, resulting in projected baselines (PBLs) and position angles (PAs) of 60\,m, 100\,m, and 130\,m, and 216\degr, 112\degr, and 63\degr. Due to non-optimal weather conditions, the observations were performed without the fringe tracker FINITO (Fringe-tracking Instrument of NIce and TOrino) \citep{Gai2004} and at medium spectral resolution (R$\sim$1500, $\Delta \upsilon \sim 200$ kms$^{-1}$). The spectral coverage in K band is in the range of 2.126\,$\mu$m to 2.474\,$\mu$m. The details of the observation are given in Table\,\ref{tab:obs}. 

The data were reduced using the software package {\it amdlib} v.3.0.8\footnote{The \textit{AMBER} reduction package \textit{amdlib} is available at: http://www.jmmc.fr/data\_processing\_amber.htm} \citep{Tatulli2007,Chelli2009}. A standard frame selection was performed of 20\% of the frames with the highest fringe signal-to-noise ratio (SNR). In order to calibrate the transfer function, the calibrator star HR\,2552 was used. The wavelength calibration was refined using the telluric absorption lines present in the spectra of both the target and the calibrator. In addition, the spectrum of the calibrator was used to correct the target spectrum for telluric features and instrumental response.

%

\begin{table*}
\caption{Log of the VLTI/\textit{AMBER} observations of HD\,50138 and calibrator HR\,2552.}             
\label{tab:obs}      
\centering     
\scalebox{0.94}{
\begin{tabular}{c c c c c c c c c c c}     
\hline\hline       
Object & Observation& \multicolumn{2}{c}{Time} & UT$\rm^{a}$ &Baseline& PA& Spectral&Wavelength& DIT$\rm^{c}$ &  UD\\ 
Name & Date & Start & End & Array &Range&& Mode$\rm^{b}$&  & & Diameter$\rm^{d}$\\
&&&&&(m)&(\rm$^{o}$)&&($\mu$m)& (s)&(mas) \\
\hline                    
  HD\,50138&2015 Dec. 23&05:54&06:25&UT1-UT3-UT4&100/60/130&216/112/63&MR-K&2.12-2.47&0.2& \\
  HR\,2552&2015 Dec. 23&06:25&06:30&UT1-UT3-UT4&100/60/130&216/112/63&MR-K&2.12-2.47&0.2&$0.65 \pm 0.04$  \\
\hline 
\end{tabular}}
\tablefoot{
$^{a}$ Unit Telescope, 
$^{b}$ Medium spectral resolution in K-band without using the fringe tracker FINITO, 
$^{c}$ Detector integration time per interferogram, 
$^{d}$ Uniform-disc diameter derived from the software package SearchCal from Jean-Marie Mariotti Center (JMMC). }
\end{table*}

\section{Results}
\subsection{Interferometric observables}
\label{sect:observables}

Our HD\,50138 observations provide us with a spectrum, three spectrally dispersed visibilities, three differential phases, and one closure phase (Fig.\,\ref{fig:intobs}). 

The spectrum shows continuum emission, as well as \HI\ emission lines of bright \brg\ and faint high-$\textit{n}$ Pfund (from level n=24 to n=19). This is in contrast to previous studies that detected only continuum and \brg\ line emission \citep{Ellerbroek2015,Fernandes2009}. 
The \brg\ peak-to-continuum ratio is $\sim$1.5, with a full width at half maximum (FWHM) of $\sim$250\,\kms; that is, it is  marginally spectrally resolved. The profile is symmetric, in agreement with previous observations of this source at higher spectral resolution   \citep[$\Delta$v$\sim$ 25\,\kms;][]{Ellerbroek2015}. 
The Pfund emission lines are weaker and their widths are consistent with that measured in \brg, with line-to-continuum flux ratios of 5\% to 10\%. Fluxes are reported in Table \ref{tab:fluxes}. 

Visibility values of the continuum are close to zero, indicating that it is spatially resolved. Low visibility values of $\sim$0.24, $\sim$0.09, and $\sim$0.13 are measured for the PBLs 60\,m, 100\,m, and 130\,m, respectively. 
Meanwhile, the visibility values of the \HI\ emission lines are higher than those of the continuum, indicating that the line-emitting region is more compact than the continuum. Correcting for the continuum contribution gives the pure line visibility (Fig.\,\ref{fig:pvis}), which provides a clearer picture of just how compact the \HI\ emitting region may be. 

A strong differential phase signature is detected across the \brg\ line in all three baselines, indicating a significant displacement of its photocentre with respect to the continuum. All three baselines show the familiar S-shaped curve characteristic of rotation. No differential phase signature is detected in the Pfund lines. 

Finally, a strong closure phase signature of $\sim$-25$\pm$ 5\degr\ in the continuum emission and  -12$\pm$ 3\degr\ in \brg\ emission is measured, indicating point-like asymmetry. No closure phase offset is detected in the Pfund lines. 
\subsection{Geometric fitting}
\label{fitting}
In order to estimate the size of the continuum and line-emitting regions a two-dimensional (2D) elongated Gaussian fitting model was used. Due to our limited uv coverage, we rely on literature values of an inclination angle of i=$56^o$ and a position angle of the major axis of the disc of PA=$71^o$ \citep[e.g.][]{Fernandes2011,Ellerbroek2015,Kluska2016}. These values were estimated using different interferometric instruments and better uv coverage. 
From the fitting we obtain the half width at half maximum (HWHM) of the brightness distribution. 
\subsection{Continuum emitting region}
\label{cont}

The three continuum visibilities cannot be fitted by a geometric 2D elongated Gaussian-disc model. 
The difficulty arises because the position angle dependence of the visibilities cannot be reproduced by a simple elongated brightness distribution. Instead, the continuum visibility measured on the 100\,m baseline is lower than that measured for 130\,m. 
Therefore, our continuum visibility measurements along with the large offset in the closure phase signal indicate that the geometry of the system is very complex, and may result from the contribution of several components with spatial extensions highly dependent on the position angle and baseline length of the individual observations. 

In the context of such complexity, we estimate the size of the continuum-emitting region by deriving a value for each baseline observation (Fig.\,\ref{fig:vis-gaussgauss}, left panel).  
The stellar component contribution was removed from the visibilities assuming a stellar-to-total flux ratio (in the K band at 2.16 \um) of $F_{*}/F_{tot}=0.08$ \citep{Ellerbroek2015}. The estimated sizes for the continuum-emitting region are given in Table\,\ref{tab:dpsize}. The HWHM values range from 1.8\,mas ($\sim$ 0.61 au)  for the longest (130 m) baseline to 3--3.4\,mas (1-1.15 au) for the remaining two baselines. 
It is worth noting that the sizes derived from the fitting (HWHM) of individual visibility points give us the projection of the whole brightness distribution of the object on that specific baseline and position angle, and not the object size at a specific position angle.


\subsubsection{\brg\, line-emitting region}
 \label{sect:Brg} 
The fact that the \brg\ line emission is both broad and bright allowed us to extract pure line visibilities in a total of five spectral channels across the \brg\ line.
To estimate the pure line visibilities we followed the method described in Appendix C of \citet{Weigelt2007}. 
We also corrected for intrinsic photospheric absorption lines using a B-type stellar template. 

\begin{figure*}[h]
\centering
\includegraphics[scale=0.8]{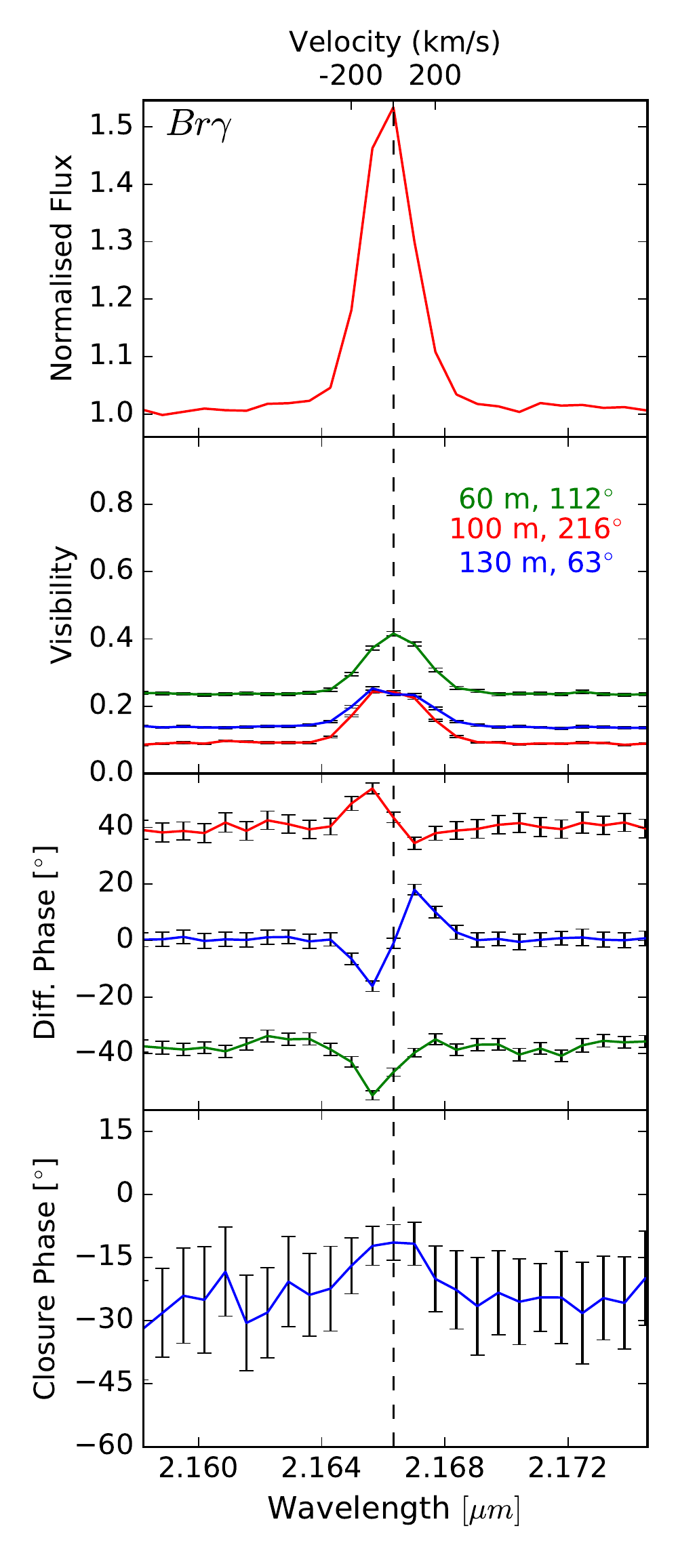}
\includegraphics[scale=0.78]{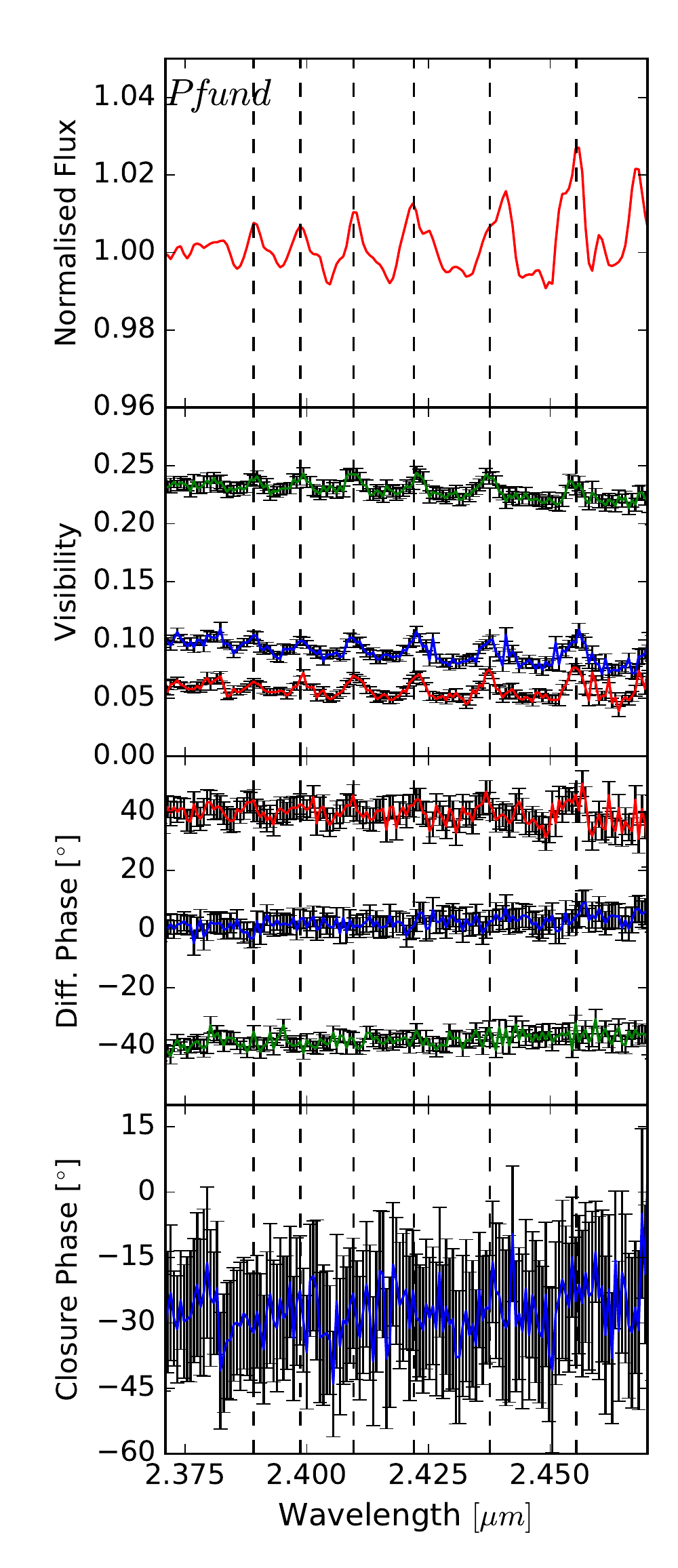}
\caption{Spectra, visibilities, differential phases and closure phase for each baseline observation, of the $Br\gamma$ line (left) and high-n Pfund lines (right). The dashed lines correspond to the systemic velocities of the lines. Flux is normalised to the continuum.}
\label{fig:intobs}
\end{figure*}

\begin{figure*}[h] 
\centering\includegraphics[scale=0.53]{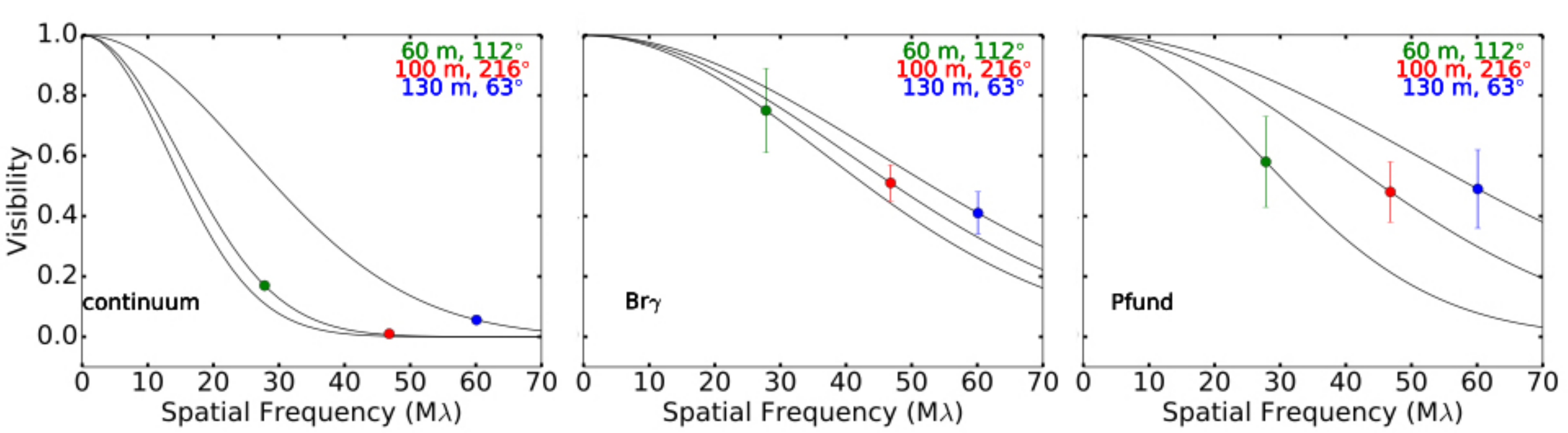}
\caption{Single 2D Gaussian fitting of visibility versus spatial frequency, for the continuum, \brg\ line and Pfund lines, respectively. Each point was fitted individually, and a size was estimated from the HWHM, Table \ref{tab:dpsize}. Above, Left panel: Continuum visibilities corrected for the stellar contribution. The errors are too small to be seen in the plot; \textit{middle panel:} \brg\ pure line visibilities; Right panel: Averaged Pfund visibilities.}
\label{fig:vis-gaussgauss}
\end{figure*}

Figure\,\ref{fig:pvis} (left panels) shows these values plotted along with the original visibilities of Fig.\,\ref{fig:intobs}, for comparison. The value of the \brg\ pure line visibility across the line profile remains constant within the error bars. We report average values of 0.75$\pm$0.14, 0.51$\pm$0.06 and 0.41$\pm$0.07 for the 60\,m, 100\,m, and 130\,m baselines respectively (Table\,\ref{tab:pvis}). We note that values decrease with increasing baseline length. 
A geometric model-fitting of a Gaussian for each  individual baseline visibility value was conducted, for ease of comparison with the continuum (Fig.\,\ref{fig:vis-gaussgauss}, centre and left panels). HWHM values ranging from $\sim$1.0 to $\sim$1.3\,mas were found, that is 0.34-0.41\,au. 
Notably, if the three \brg\ pure line visibilities  are fitted simultaneously with a 2D Gaussian, a size of 1.00 $\pm$ 0.13 mas is found, corresponding to 0.36 $\pm$ 0.05 au. This result is consistent with the range of values obtained from the individual fittings.
Similarly, correcting the closure phase for the continuum contribution reveals an offset of $\sim$+20\degr$\pm$10\degr\ (Fig.\,\ref{fig:pvis}, left panels: the pure line closure phases have been computed  by using the method also described in Appendix C of \citet{Weigelt2007}). This offset is opposite in direction to that derived for the continuum. 

\subsubsection{High-$\textit{n}$ Pfund line-emitting region}
\label{sect:pfund}

\begin{figure*}[h]
\centering
\includegraphics[scale = 0.8]{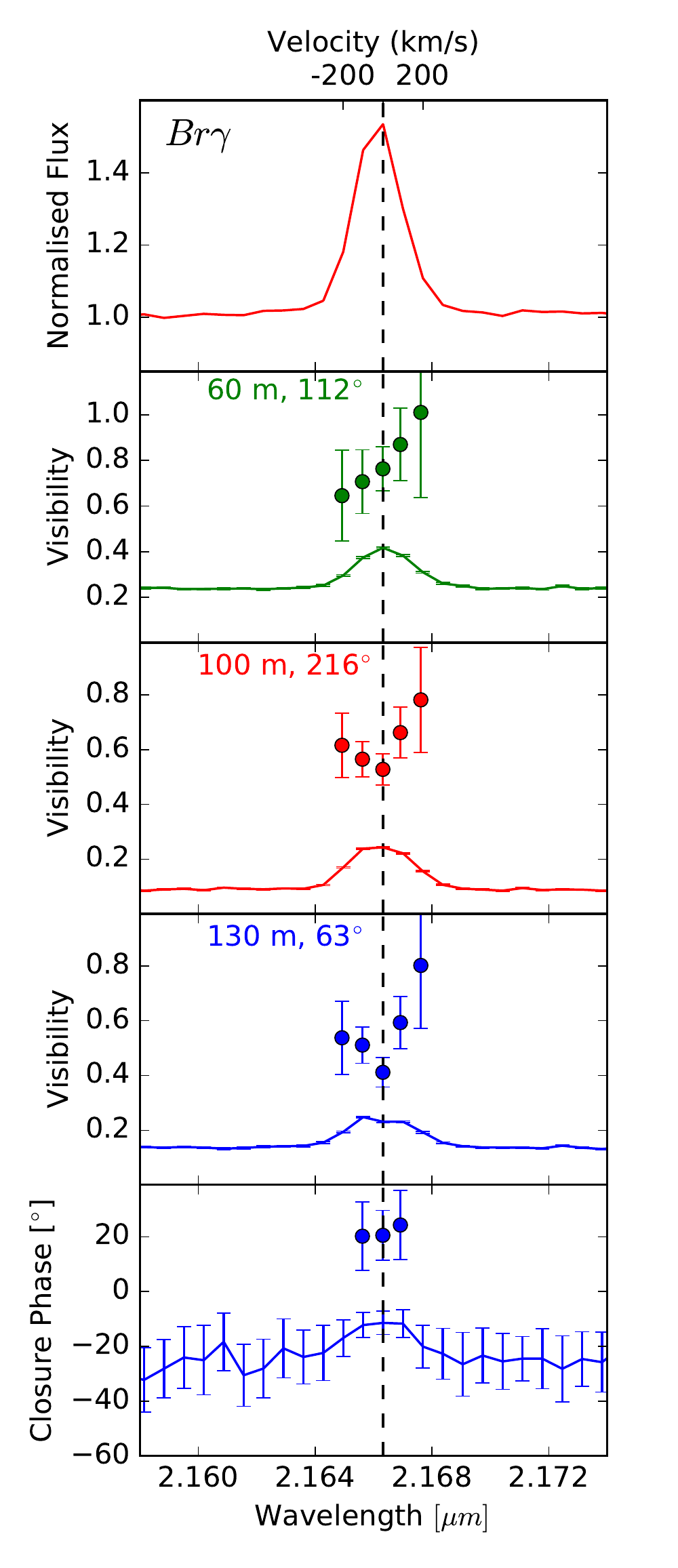}
\includegraphics[scale = 0.78]{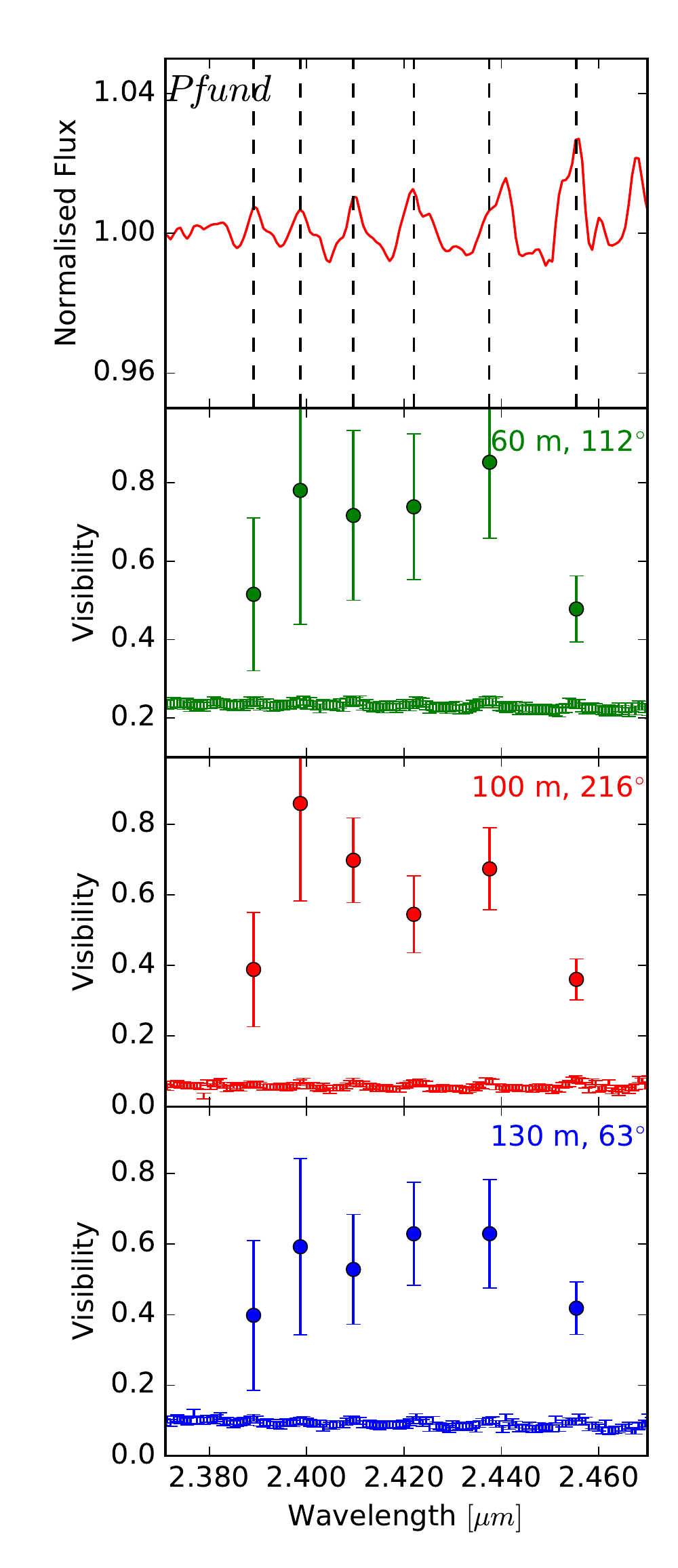}
\caption{Spectrum, and pure line visibilities for each baseline observation of the $Br\gamma$ line (left) and high-n Pfund lines (right). The dashed lines correspond to the systemic velocities of the lines. For clarity, the dashed lines were removed for the Pfund visibility plots. The visibilities of Figure\,\ref{fig:intobs} are overplotted, for comparison. The continuum-corrected closure phase for the \brg\,line is also plotted, and accompanied by an overplot of the closure phase of Fig.\,\ref{fig:intobs}. No offset in closure phase for Pfund was detected in Fig.\,\ref{fig:intobs}, and so no plot is made here of the continuum-corrected closure phase for Pfund.} 
\label{fig:pvis}
\end{figure*}

The derived pure line visibilities for the Pfund emission are shown in Fig.\,\ref{fig:pvis} (right panel).
Values across all lines are consistent within errors. Therefore, in order to achieve a higher accuracy in estimating the size of the Pfund emitting region, we calculate the weighted-mean of the six Pfund visibilities for a given baseline. The resulting three Pfund pure line visibilities (Table\, \ref{tab:pvis}) then allowed a Gaussian fit to estimate the size of the emitting region. We find a Pfund line-emitting region of size r$\sim$0.3--0.57\,au. This is very close to our value for the size of the \brg\ line-emitting region of r$\sim$0.34--0.41\,au (Sect.\,\ref{sect:Brg}).


\begin{table}[h]
\caption{Correlated fluxes of \HI\,~emission lines.}             
\label{tab:fluxes}      
\centering                          
\begin{tabular}{l c }        
\hline\hline                 
Emission line &Flux  \\ 
&(10$^{-13}$ ergs s$^{-1}$cm$^{-2}$) \\
\hline                        
   \brg & 81$\pm$1.20  \\      
   Pfund24-5 & 2.6$\pm$0.76 \\
   Pfund23-5& 2.1$\pm$0.63  \\
   Pfund22-5& 2.8$\pm$0.84  \\
   Pfund21-5& 3.2$\pm$0.96 \\
   Pfund20-5& 3.2$\pm$0.96  \\
   Pfund19-5& 6.7$\pm$0.80  \\
   
\hline                                   
\end{tabular} 
\end{table}


\begin{table}[h]
\caption{Continuum and pure line visibilities.}             
\label{tab:pvis}      
\centering                          
\begin{tabular}{c c c c}        
\hline\hline                 
Baseline& Continuum & \brg & Pfund\\    
(m)&&&\\
\hline                        
   60 &0.239$\pm 0.004$& 0.75$\pm 0.14$ & 0.58$\pm 0.15$  \\
   101 &0.092$\pm 0.002$& 0.51$\pm 0.06$ & 0.48$\pm 0.10$ \\
   130&0.135$\pm 0.003$ & 0.41 $\pm 0.07$& 0.49 $\pm 0.13$  \\      

\hline                                   
\end{tabular} 
\tablefoot{For the \brg\, emission the visibility of the peak of the line is recorded, while in the case of the Pfund lines a weighted mean average value of the lines is reported. The \brg\, and Pfund pure line visibility values of this table are shown in Figure \ref{fig:pvis}.}
\end{table}

\begin{table*}
\caption{Sizes (HWHM) of the emitting regions.}             
\label{tab:dpsize}      
\centering     
\begin{tabular}{c c c c c c c}     
\hline\hline       
Baseline& \multicolumn{2}{c}{Continuum}& \multicolumn{2}{c}{\brg} &  \multicolumn{2}{c}{Pfund}\\ 
m & mas & au & mas &au&mas&au\\
\hline  
60 &3.00$\pm 0.09$&1.03$\pm 0.03$&1.22$\pm 0.23$& 0.41$\pm 0.08$ &  1.7$\pm 0.41$&0.57$\pm 0.14$   \\  
101 &3.40$\pm 0.15$&1.15$\pm 0.05$&1.11$\pm 0.12$& 0.38$\pm 0.04$ &1.16$\pm 0.15$&  0.37$\pm 0.06$  \\
130 &1.80$\pm 0.02$& 0.61 $\pm 0.006$&1.00$\pm 0.18$& 0.34$\pm 0.06$ &0.89$\pm 0.19$&0.30$\pm 0.06$ \\ 

\hline                  
\end{tabular}
\tablefoot{An inclination of $56 \pm 4^o$ and a position angle of the major axis of the disc of $71\pm7^o$ \citep{Fernandes2011} were inputs in the 2D elongated Gaussian model. The continuum is corrected for the stellar contribution.}
\end{table*}

%




\section{Discussion}
\label{sect:discussion}

\subsection{Morphology and kinematics} 

The picture emerging from our observations is of a highly complex and asymmetric circumstellar environment in the immediate vicinity of \HD\ . 

		Our continuum visibilities for the longest baseline agree well with reports of an inclined circumstellar disc of radius $\sim$2\,mas, based on multi-epoch VLTI/{\it PIONIER} interferometric observations of this source in the H band \citep{Kluska2016,Lazareff2016}. However, our shorter baseline visibilities give a continuum size almost twice as large. Interestingly, this larger size is very similar to that derived from K-band \textit{AMBER} interferometric studies of this source giving a continuum size (after correcting for the stellar component) of $\sim$3.4\,mas \citep{Ellerbroek2015}. It should be noted, however, that the best-fit models of \citet{Lazareff2016} and \cite{Ellerbroek2015} utilize a greater number of shorter than longer baselines and so are unevenly weighted. Overall, it seems that the different sizes derived for the continuum emission region in the various studies point to a complex and variable geometry of the \HD\ circumstellar environment. 


The differential phase signature across the \brg\ line presents a clear S-shape, which indicates that the gas is rotating. In the case of a rotating disc, which is not fully resolved, the differential phase signature acquires an S-shape profile with maximum amplitude measured when using baselines aligned with the disc major axis. To test if this rotating gas originates close to the disc surface (either from the disc surface itself, or at the base of a rotating disc wind), we examine the \brg\ profile for consistency with a disc in Keplerian rotation. Assuming a distance to \HD\ of 340\,pc, the \brg\ emission originates in a region which extends out to about 0.4\,au from the star. At this distance, the Keplerian velocity of the disc is $\sim$100\,\kms\ (assuming $i$=54\degr). We find that our measured HWHM of the \brg\ line is roughly this value, and thus supports an origin from rotating gas near or on the disc surface. Indeed, a model of the disc in Keplerian rotation provided a good fit to previously reported high-resolution \textit{AMBER} data \citep{Ellerbroek2015}. However, such a model alone fails in reproducing some of our interferometric observables, namely, a large offset in the continuum closure phase signal \citep[not detected in][]{Ellerbroek2015}, as well as the offset in the \brg\ closure phase signal detected here for the first time. 

The large offset of $\sim$-25\degr\ in the continuum closure phase signal is evidence of a significant asymmetry in the circumstellar structure. Previous reports of this asymmetry interpreted it as an unevenly illuminated asymmetric disc with the eastern side much brighter than the western side, based on multi-epoch VLTI/{\it PIONIER} H-band observations  \citep{Lazareff2016}. Other studies modelled the asymmetry as a bright spot that moves around the star, based on two-epoch reconstructed images \citep{Kluska2016}. Time-variable asymmetries are indeed common in B[e] stars \citep{Okazaki1991,Carciofi2009,Schaefer2010}. Long term variability (i.e. on decade timescales) is often explained by the development of a one-armed spiral superimposed on a circumstellar disc \citep{Okazaki1991,Papaloizou1992}. This spiral density wave precesses with the disc rotation, producing a difference in brightness between the two sides of the disc. Such a scenario could account for the large offset in the continuum closure phase signal, along with the aforementioned S-shape in the \brg\ differential phase signature. Indeed, this explanation has been proposed for other Be stars \citep{Schaefer2010,Oktariani2016}.

However, there is also an offset of $\sim$+20\degr\ in the \brg\ closure phase signal which, similarly, indicates an asymmetry. Confusingly, this asymmetry is in a direction opposite to that of the continuum asymmetry. 
A possible scenario could be the presence of an outflow in which one lobe is much brighter than the other. This could be for instance produced by the presence of a hot spot in the stellar surface which leads to a preferred ejection direction \citep{Humphreys2002}. Other possible scenarios include the presence of a binary emitting in \brg\ line, or an axisymmetric wind in which part of the emission is blocked by the (optically thick) disc. None of these scenarios would, however, reproduce the S-shaped \brg\ line differential phase. Asymmetries of the continuum and the \brg\, line-emitting regions have been reported in evolved stars \citep[see, e.g.][]{Meilland2013,Driebe2009}. The similarity of our results with these previous reports seems to imply that \HD\ is an evolved star.


Our interferometric observables make a clear case for the complexity of the continuum and \brg\ line circumstellar environment of \HD\, which is difficult to interpret with our single epoch observations. From our analysis, we see that no single component can simultaneously explain the two closure phases and the S-shape of the \brg\ profile, but instead we find we must evoke multiple components. Providing meaningful constraints on this complex system demands improved coverage, both in the \textit{uv}-plane and temporally, as well as higher spectral resolution. 

\subsection{\HI\ physical conditions} 

\begin{figure}[h]
\centering\includegraphics[width=\columnwidth]{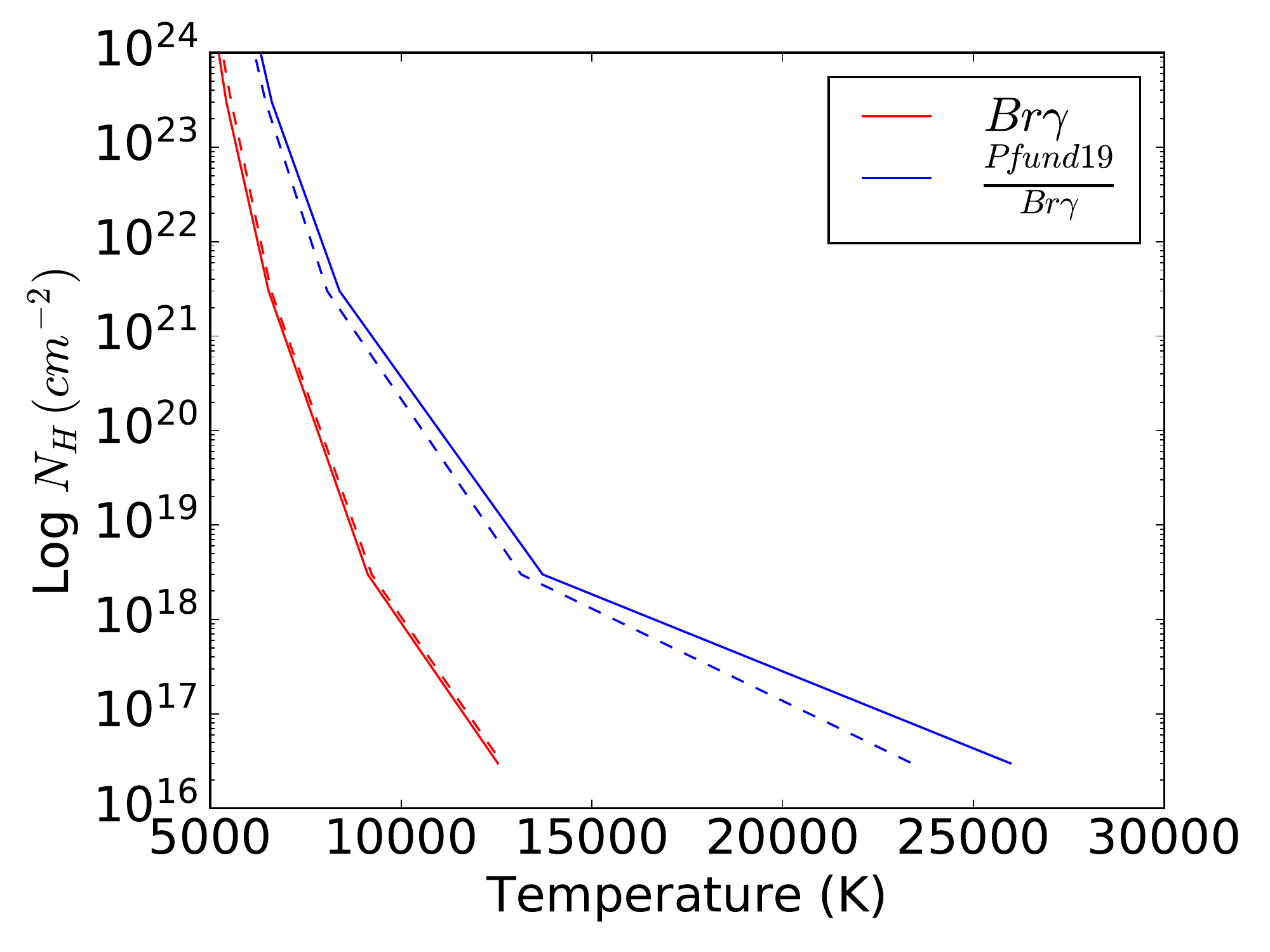}
\caption{Locus of the column density of neutral hydrogen ($N_{H}$) and temperature (T) that best reproduce the fluxes of \brg\ and the Pfund 19 (Table \ref{tab:fluxes}). The red solid line is for \brg\, while the blue solid line represents the ratio of the Pfund 19 over the \brg. The dashed lines are the 3 $\sigma$ uncertainties plotted only from one side. The ratios of the other Pfund lines are not plotted because they are very similar to that for Pfund 19.}
\label{fig:locus_whole_disc}
\end{figure}

The simultaneous detection of the \brg\ and Pfund lines allows us to provide additional information on some of the physical properties of the emitting region. First of all, the combination of the flux of the \brg\ line and the size of the emitting region derived from our interferometric data shows that the \brg\ line is unlikely to be optically thick. Assuming an area of the emitting region of $1.12\times 10^{26}$\,cm$^2$ and a correlated flux in the line of $\sim 8\times 10^{-12}$\,erg\,s$^{-1}$\,cm$^{-2}$, a simple black-body estimate gives a temperature value of about 2200\,K. 
At this temperature, assuming an LTE level population, the \brg\ line will be optically thick only for unrealistically large column densities (i.e. $N_H \simgreat 10^{31}$ cm$^{-2}$). 
At the same time, the ratio of the Pfund line fluxes over \brg\ ($\sim 5-10$\%) seem to rule out the lines forming in a recombination cascade. Namely, in Case B recombination \citep{Baker1938,Storey1995} 
this ratio is never larger than 1-2\% even for temperatures as high as 30\,000\,K. 
We adopt a very simple model for the \HI\ emitting region, namely an isothermal, constant density slab. Assuming that the level populations are in LTE and a line width of 250\,\kms, we compute a grid of solutions on the ($T,N_H$) plane to be compared with the two 
constraints provided by our observations: the \brg\ flux and the flux ratios of Pfund lines to \brg. These values depend on the temperature $T$, column density $N_H$, and the size of the emitting region ($r\sim$0.4\,au) as derived from our interferometric data. The locus of the points that reproduce the observed \brg\  flux is shown as a red solid line in Fig.\,\ref{fig:locus_whole_disc}. 
The coloured solid curves represent the locus of the ($T,N_H$) points that reproduce the observed values of the ratio of the Pfund  over  \brg\ fluxes, assuming that the emitting region is the same for all the lines. 
From Fig.\,\ref{fig:locus_whole_disc} it is clear that there is no set of (T,N$_H$) values that simultaneously reproduce the \brg\  flux and the Pfund to \brg\ line ratios within the 3$\sigma$ uncertainties of our measurements, with slightly better agreement  for low $T$ and high $N_H$. 
%
For fixed N$_H$, the region emitting the Pfund lines must be slightly hotter than that emitting the \brg\ line. 

A possible  explanation of this result is that the  Pfund lines form in a region where both $T$ and the gas density vary along the line of sight, as expected in an outflow where the density decreases and the temperature increases moving away from the star. The similarity of the projected sizes in a system with the inclination of \HD\ suggests that the outflow probably has a relatively wide opening angle. 

\cite{Kraus2012} present \textit{AMBER} interferometric observations of the B[e] star $\zeta$ Tau (\Tstar =19370K, Log \Lstar/\Lsun=3.9), which include \brg\ and 4 Pfund lines from  n=19-5 to n=22-5.  In $\zeta$ Tau  the Pfund lines come from a smaller region than \brg\ by a factor of approximately two and they are brighter relative to \brg\ than in \HD. Those authors modelled the Pfund lines using a simple, LTE configuration with power-law radial gradients both in temperature and column density, following \citet{Stee1994} and \citet{Carciofi2008}. 
If we apply a similar model to \HD\,  at 0.4 au from the star we expect $T\sim 4000$ K and N$_H \sim 10^{21}$\,$cm^2$, which cannot explain our observations.
Detailed models, which include non-LTE effects \citep[e.g.,][]{Carciofi2008}, are required before any firm conclusion can be reached.

\section{Conclusions}

We report observations of the Herbig B[e] star, \HD, using VLTI/\textit{AMBER} interferometric observations in the K band at medium spectral resolution. Our observations allowed us to map the size of the continuum emission and of the \brg\ and high-n Pfund lines. The latter is detected for the first time. Our results reveal that the circumstellar environment is very complex. 

Morphologically, the continuum was found to be well resolved, but could not be reproduced using a symmetric Gaussian disc model for the brightness distribution. Nevertheless, a projected size of 0.6-1 au was estimated. The \HI\ lines are more compact than the continuum-emitting region; a projected size of $\sim$ 0.4 au was estimated. Both continuum and \brg\ line were found to trace an asymmetric origin but, confusingly, with opposing directions of asymmetries. Various possible scenarios are discussed, but nothing could be concluded without better uv-plane sampling. 

Kinematically, the \brg\ line presented a signature consistent with Keplerian rotation, and supporting the case for an origin close to the disc surface (either in the disc surface itself, or at the base of a rotating disc wind). No such signature was detected for the Pfund lines, possibly due to the lower SNR. 

The \HI\ physical conditions were investigated by applying a simple model to the \HI\ line fluxes and their sizes. We conclude that the \brg\ line is optically thin, and traces a wind with a wide opening angle.

Although we cannot exclude the possibility that \HD\ is a pre-main sequence star, the similarity of our results with studies of evolved stars favours an evolved status also for \HD.

\begin{acknowledgements}
M.K. is funded by the Irish Research Council (IRC), grant GOIPG/2016/769 and SFI grant 13/ERC/12907. R.G.L has received funding from the European Union's Horizon 2020 research and innovation programme under the Marie Sk\l{}odowska-Curie Grant (agreement No.\ 706320). A.C.G. and T.P.R. have received funding from the European Research Council (ERC) under the European Union's Horizon 2020 research and innovation programme (grant agreement No.\ 743029). JSB acknowledges the support from the ESO Fellowship programme.
\end{acknowledgements}


\bibliographystyle{aa}
\bibliography{references.bib}{}

\begin{thebibliography}{39}
\expandafter\ifx\csname natexlab\endcsname\relax\def\natexlab#1{#1}\fi

\bibitem[{{Andrillat} \& {Houziaux}(1991)}]{Andrillat1991}
{Andrillat}, Y. \& {Houziaux}, L. 1991, \iaucirc, 5164

\bibitem[{{Baker} \& {Menzel}(1938)}]{Baker1938}
{Baker}, J.~G. \& {Menzel}, D.~H. 1938, \apj, 88, 52

\bibitem[{{Borges Fernandes} {et~al.}(2009){Borges Fernandes}, {Kraus},
  {Chesneau}, {Domiciano de Souza}, {de Ara{\'u}jo}, {Stee}, \&
  {Meilland}}]{Fernandes2009}
{Borges Fernandes}, M., {Kraus}, M., {Chesneau}, O., {et~al.} 2009, \aap, 508,
  309

\bibitem[{{Borges Fernandes} {et~al.}(2011){Borges Fernandes}, {Meilland},
  {Bendjoya}, {Domiciano de Souza}, {Niccolini}, {Chesneau}, {Millour},
  {Spang}, {Stee}, \& {Kraus}}]{Fernandes2011}
{Borges Fernandes}, M., {Meilland}, A., {Bendjoya}, P., {et~al.} 2011, \aap,
  528, A20

\bibitem[{{Calvet} {et~al.}(2004){Calvet}, {Muzerolle}, {Brice{\~n}o},
  {Hern{\'a}ndez}, {Hartmann}, {Saucedo}, \& {Gordon}}]{Calvet2004}
{Calvet}, N., {Muzerolle}, J., {Brice{\~n}o}, C., {et~al.} 2004, \aj, 128, 1294

\bibitem[{{Caratti o Garatti} {et~al.}(2015){Caratti o Garatti}, {Tambovtseva},
  {Garcia Lopez}, {Kraus}, {Schertl}, {Grinin}, {Weigelt}, {Hofmann}, {Massi},
  {Lagarde}, {Vannier}, \& {Malbet}}]{Alessio2015}
{Caratti o Garatti}, A., {Tambovtseva}, L.~V., {Garcia Lopez}, R., {et~al.}
  2015, \aap, 582, A44

\bibitem[{{Carciofi} \& {Bjorkman}(2008)}]{Carciofi2008}
{Carciofi}, A.~C. \& {Bjorkman}, J.~E. 2008, \apj, 684, 1374

\bibitem[{{Carciofi} {et~al.}(2009){Carciofi}, {Okazaki}, {Le Bouquin}, {{\v
  S}tefl}, {Rivinius}, {Baade}, {Bjorkman}, \& {Hummel}}]{Carciofi2009}
{Carciofi}, A.~C., {Okazaki}, A.~T., {Le Bouquin}, J.-B., {et~al.} 2009, \aap,
  504, 915

\bibitem[{{Chelli} {et~al.}(2009){Chelli}, {Utrera}, \& {Duvert}}]{Chelli2009}
{Chelli}, A., {Utrera}, O.~H., \& {Duvert}, G. 2009, \aap, 502, 705

\bibitem[{{Driebe} {et~al.}(2009){Driebe}, {Groh}, {Hofmann}, {Ohnaka},
  {Kraus}, {Millour}, {Murakawa}, {Schertl}, {Weigelt}, {Petrov}, {Wittkowski},
  {Hummel}, {Le Bouquin}, {Merand}, {Sch{\"o}ller}, {Massi}, {Stee}, \&
  {Tatulli}}]{Driebe2009}
{Driebe}, T., {Groh}, J.~H., {Hofmann}, K.-H., {et~al.} 2009, \aap, 507, 301

\bibitem[{{Ellerbroek} {et~al.}(2015){Ellerbroek}, {Benisty}, {Kraus},
  {Perraut}, {Kluska}, {le Bouquin}, {Borges Fernandes}, {Domiciano de Souza},
  {Maaskant}, {Kaper}, {Tramper}, {Mourard}, {Tallon-Bosc}, {ten Brummelaar},
  {Sitko}, {Lynch}, \& {Russell}}]{Ellerbroek2015}
{Ellerbroek}, L.~E., {Benisty}, M., {Kraus}, S., {et~al.} 2015, \aap, 573, A77

\bibitem[{{Gai} {et~al.}(2004){Gai}, {Corcione}, \& {Massone}}]{Gai2004}
{Gai}, M., {Corcione}, L., \& {Massone}, G. 2004, in Astrophysics and Space
  Science Library, Vol. 300, Scientific Detectors for Astronomy, The Beginning
  of a New Era, ed. P.~{Amico}, J.~W. {Beletic}, \& J.~E. {Belectic}, 341--344

\bibitem[{{Garcia Lopez} {et~al.}(2016){Garcia Lopez}, {Kurosawa}, {Caratti o
  Garatti}, {Kreplin}, {Weigelt}, {Tambovtseva}, {Grinin}, \&
  {Ray}}]{Rebeca2016}
{Garcia Lopez}, R., {Kurosawa}, R., {Caratti o Garatti}, A., {et~al.} 2016,
  \mnras, 456, 156

\bibitem[{{Garcia Lopez} {et~al.}(2015){Garcia Lopez}, {Tambovtseva},
  {Schertl}, {Grinin}, {Hofmann}, {Weigelt}, \& {Caratti o
  Garatti}}]{Rebeca2015}
{Garcia Lopez}, R., {Tambovtseva}, L.~V., {Schertl}, D., {et~al.} 2015, \aap,
  576, A84

\bibitem[{{Gray} \& {Corbally}(1998)}]{Gray1998}
{Gray}, R.~O. \& {Corbally}, C.~J. 1998, \aj, 116, 2530

\bibitem[{{Harrington} \& {Kuhn}(2007)}]{Harrington2007}
{Harrington}, D.~M. \& {Kuhn}, J.~R. 2007, \apjl, 667, L89

\bibitem[{{Harrington} \& {Kuhn}(2009)}]{Harrington2009}
{Harrington}, D.~M. \& {Kuhn}, J.~R. 2009, \apjs, 180, 138

\bibitem[{{Humphreys} {et~al.}(2002){Humphreys}, {Davidson}, \&
  {Smith}}]{Humphreys2002}
{Humphreys}, R.~M., {Davidson}, K., \& {Smith}, N. 2002, \aj, 124, 1026

\bibitem[{{Hutsemekers}(1985)}]{Hutsemekers1985}
{Hutsemekers}, D. 1985, \aaps, 60, 373

\bibitem[{{Kluska} {et~al.}(2016){Kluska}, {Benisty}, {Soulez}, {Berger}, {Le
  Bouquin}, {Malbet}, {Lazareff}, \& {Thi{\'e}baut}}]{Kluska2016}
{Kluska}, J., {Benisty}, M., {Soulez}, F., {et~al.} 2016, \aap, 591, A82

\bibitem[{{Kraus} {et~al.}(2008){Kraus}, {Hofmann}, {Benisty}, {Berger},
  {Chesneau}, {Isella}, {Malbet}, {Meilland}, {Nardetto}, {Natta}, {Preibisch},
  {Schertl}, {Smith}, {Stee}, {Tatulli}, {Testi}, \& {Weigelt}}]{Kraus2008}
{Kraus}, S., {Hofmann}, K.-H., {Benisty}, M., {et~al.} 2008, \aap, 489, 1157

\bibitem[{{Kraus} {et~al.}(2012){Kraus}, {Monnier}, {Che}, {Schaefer},
  {Touhami}, {Gies}, {Aufdenberg}, {Baron}, {Thureau}, {ten Brummelaar},
  {McAlister}, {Turner}, {Sturmann}, \& {Sturmann}}]{Kraus2012}
{Kraus}, S., {Monnier}, J.~D., {Che}, X., {et~al.} 2012, \apj, 744, 19

\bibitem[{{Lazareff} {et~al.}(2016){Lazareff}, {Berger}, {Kluska}, {Le
  Bouquin}, {Benisty}, {Malbet}, {Koen}, {Pinte}, {Thi}, {Absil}, {Baron}, \&
  {et al.}}]{Lazareff2016}
{Lazareff}, {Berger}, {Kluska}, {et~al.} 2016, \aap

\bibitem[{{Lindegren} {et~al.}(2016){Lindegren}, {Lammers}, {Bastian},
  {Hern{\'a}ndez}, {Klioner}, {Hobbs}, {Bombrun}, {Michalik}, {Ramos-Lerate},
  {Butkevich}, {Comoretto}, {Joliet}, {Holl}, {Hutton}, {Parsons},
  {Steidelm{\"u}ller}, {Abbas}, {Altmann}, {Andrei}, {Anton}, {Bach},
  {Barache}, {Becciani}, {Berthier}, {Bianchi}, {Biermann}, {Bouquillon},
  {Bourda}, {Br{\"u}semeister}, {Bucciarelli}, {Busonero}, {Carlucci},
  {Casta{\~n}eda}, {Charlot}, {Clotet}, {Crosta}, {Davidson}, {de Felice},
  {Drimmel}, {Fabricius}, {Fienga}, {Figueras}, {Fraile}, {Gai}, {Garralda},
  {Geyer}, {Gonz{\'a}lez-Vidal}, {Guerra}, {Hambly}, {Hauser}, {Jordan},
  {Lattanzi}, {Lenhardt}, {Liao}, {L{\"o}ffler}, {McMillan}, {Mignard}, {Mora},
  {Morbidelli}, {Portell}, {Riva}, {Sarasso}, {Serraller}, {Siddiqui}, {Smart},
  {Spagna}, {Stampa}, {Steele}, {Taris}, {Torra}, {van Reeven}, {Vecchiato},
  {Zschocke}, {de Bruijne}, {Gracia}, {Raison}, {Lister}, {Marchant},
  {Messineo}, {Soffel}, {Osorio}, {de Torres}, \& {O'Mullane}}]{GAIA2016}
{Lindegren}, L., {Lammers}, U., {Bastian}, U., {et~al.} 2016, \aap, 595, A4

\bibitem[{{Maddalena} {et~al.}(1986){Maddalena}, {Morris}, {Moscowitz}, \&
  {Thaddeus}}]{Maddalena1986}
{Maddalena}, R.~J., {Morris}, M., {Moscowitz}, J., \& {Thaddeus}, P. 1986,
  \apj, 303, 375

\bibitem[{{Meilland} {et~al.}(2013){Meilland}, {Stee}, {Spang}, {Malbet},
  {Massi}, \& {Schertl}}]{Meilland2013}
{Meilland}, A., {Stee}, P., {Spang}, A., {et~al.} 2013, \aap, 550, L5

\bibitem[{{Morrison} \& {Beaver}(1995)}]{Morrison1995}
{Morrison}, N.~D. \& {Beaver}, M. 1995, in Bulletin of the American
  Astronomical Society, Vol.~27, American Astronomical Society Meeting
  Abstracts \#186, 825

\bibitem[{{Muzerolle} {et~al.}(1998){Muzerolle}, {Hartmann}, \&
  {Calvet}}]{Muzerolle1998}
{Muzerolle}, J., {Hartmann}, L., \& {Calvet}, N. 1998, \aj, 116, 2965

\bibitem[{{Okazaki}(1991)}]{Okazaki1991}
{Okazaki}, A.~T. 1991, \pasj, 43, 75

\bibitem[{{Oktariani} {et~al.}(2016){Oktariani}, {Okazaki}, {Kunjaya}, \&
  {Aprilia}}]{Oktariani2016}
{Oktariani}, F., {Okazaki}, A.~T., {Kunjaya}, C., \& {Aprilia}. 2016, \mnras,
  459, 4440

\bibitem[{{Oudmaijer} \& {Drew}(1999)}]{Oudmaijer1999}
{Oudmaijer}, R.~D. \& {Drew}, J.~E. 1999, \mnras, 305, 166

\bibitem[{{Papaloizou} {et~al.}(1992){Papaloizou}, {Savonije}, \&
  {Henrichs}}]{Papaloizou1992}
{Papaloizou}, J.~C., {Savonije}, G.~J., \& {Henrichs}, H.~F. 1992, \aap, 265,
  L45

\bibitem[{{Petrov} {et~al.}(2007){Petrov}, {Malbet}, {Weigelt}, {Antonelli},
  {Beckmann}, {Bresson}, {Chelli}, {Dugu{\'e}}, {Duvert}, {Gennari},
  {Gl{\"u}ck}, {Kern}, {Lagarde}, {Le Coarer}, {Lisi}, {Millour}, {Perraut},
  {Puget}, {Rantakyr{\"o}}, {Robbe-Dubois}, {Roussel}, {Salinari}, {Tatulli},
  {Zins}, {Accardo}, {Acke}, {Agabi}, {Altariba}, {Arezki}, {Aristidi},
  {Baffa}, {Behrend}, {Bl{\"o}cker}, {Bonhomme}, {Busoni}, {Cassaing},
  {Clausse}, {Colin}, {Connot}, {Delboulb{\'e}}, {Domiciano de Souza},
  {Driebe}, {Feautrier}, {Ferruzzi}, {Forveille}, {Fossat}, {Foy},
  {Fraix-Burnet}, {Gallardo}, {Giani}, {Gil}, {Glentzlin}, {Heiden},
  {Heininger}, {Hernandez Utrera}, {Hofmann}, {Kamm}, {Kiekebusch}, {Kraus},
  {Le Contel}, {Le Contel}, {Lesourd}, {Lopez}, {Lopez}, {Magnard}, {Marconi},
  {Mars}, {Martinot-Lagarde}, {Mathias}, {M{\`e}ge}, {Monin}, {Mouillet},
  {Mourard}, {Nussbaum}, {Ohnaka}, {Pacheco}, {Perrier}, {Rabbia}, {Rebattu},
  {Reynaud}, {Richichi}, {Robini}, {Sacchettini}, {Schertl}, {Sch{\"o}ller},
  {Solscheid}, {Spang}, {Stee}, {Stefanini}, {Tallon}, {Tallon-Bosc}, {Tasso},
  {Testi}, {Vakili}, {von der L{\"u}he}, {Valtier}, {Vannier}, \&
  {Ventura}}]{Petrov2007}
{Petrov}, R.~G., {Malbet}, F., {Weigelt}, G., {et~al.} 2007, \aap, 464, 1

\bibitem[{{Schaefer} {et~al.}(2010){Schaefer}, {Gies}, {Monnier}, {Richardson},
  {Touhami}, {Zhao}, {Che}, {Pedretti}, {Thureau}, {ten Brummelaar},
  {McAlister}, {Ridgway}, {Sturmann}, {Sturmann}, {Turner}, {Farrington}, \&
  {Goldfinger}}]{Schaefer2010}
{Schaefer}, G.~H., {Gies}, D.~R., {Monnier}, J.~D., {et~al.} 2010, \aj, 140,
  1838

\bibitem[{{Stee} \& {de Araujo}(1994)}]{Stee1994}
{Stee}, P. \& {de Araujo}, F.~X. 1994, \aap, 292, 221

\bibitem[{{Storey} \& {Hummer}(1995)}]{Storey1995}
{Storey}, P.~J. \& {Hummer}, D.~G. 1995, \mnras, 272, 41

\bibitem[{{Tatulli} {et~al.}(2007){Tatulli}, {Isella}, {Natta}, {Testi},
  {Marconi}, {Malbet}, {Stee}, {Petrov}, {Millour}, {Chelli}, {Duvert},
  {Antonelli}, {Beckmann}, {Bresson}, {Dugu{\'e}}, {Gennari}, {Gl{\"u}ck},
  {Kern}, {Lagarde}, {Le Coarer}, {Lisi}, {Perraut}, {Puget}, {Rantakyr{\"o}},
  {Robbe-Dubois}, {Roussel}, {Weigelt}, {Zins}, {Accardo}, {Acke}, {Agabi},
  {Altariba}, {Arezki}, {Aristidi}, {Baffa}, {Behrend}, {Bl{\"o}cker},
  {Bonhomme}, {Busoni}, {Cassaing}, {Clausse}, {Colin}, {Connot},
  {Delboulb{\'e}}, {Domiciano de Souza}, {Driebe}, {Feautrier}, {Ferruzzi},
  {Forveille}, {Fossat}, {Foy}, {Fraix-Burnet}, {Gallardo}, {Giani}, {Gil},
  {Glentzlin}, {Heiden}, {Heininger}, {Hernandez Utrera}, {Hofmann}, {Kamm},
  {Kiekebusch}, {Kraus}, {Le Contel}, {Le Contel}, {Lesourd}, {Lopez}, {Lopez},
  {Magnard}, {Mars}, {Martinot-Lagarde}, {Mathias}, {M{\`e}ge}, {Monin},
  {Mouillet}, {Mourard}, {Nussbaum}, {Ohnaka}, {Pacheco}, {Perrier}, {Rabbia},
  {Rebattu}, {Reynaud}, {Richichi}, {Robini}, {Sacchettini}, {Schertl},
  {Sch{\"o}ller}, {Solscheid}, {Spang}, {Stefanini}, {Tallon}, {Tallon-Bosc},
  {Tasso}, {Vakili}, {von der L{\"u}he}, {Valtier}, {Vannier}, \&
  {Ventura}}]{Tatulli2007}
{Tatulli}, E., {Isella}, A., {Natta}, A., {et~al.} 2007, \aap, 464, 55

\bibitem[{{Weigelt} {et~al.}(2011){Weigelt}, {Grinin}, {Groh}, {Hofmann},
  {Kraus}, {Miroshnichenko}, {Schertl}, {Tambovtseva}, {Benisty}, {Driebe},
  {Lagarde}, {Malbet}, {Meilland}, {Petrov}, \& {Tatulli}}]{Weigelt2011}
{Weigelt}, G., {Grinin}, V.~P., {Groh}, J.~H., {et~al.} 2011, \aap, 527, A103

\bibitem[{{Weigelt} {et~al.}(2007){Weigelt}, {Kraus}, {Driebe}, {Petrov},
  {Hofmann}, {Millour}, {Chesneau}, {Schertl}, {Malbet}, {Hillier}, {Gull},
  {Davidson}, {Domiciano de Souza}, {Antonelli}, {Beckmann}, {Bresson},
  {Chelli}, {Dugu{\'e}}, {Duvert}, {Gennari}, {Gl{\"u}ck}, {Kern}, {Lagarde},
  {Le Coarer}, {Lisi}, {Perraut}, {Puget}, {Rantakyr{\"o}}, {Robbe-Dubois},
  {Roussel}, {Tatulli}, {Zins}, {Accardo}, {Acke}, {Agabi}, {Altariba},
  {Arezki}, {Aristidi}, {Baffa}, {Behrend}, {Bl{\"o}cker}, {Bonhomme},
  {Busoni}, {Cassaing}, {Clausse}, {Colin}, {Connot}, {Delboulb{\'e}},
  {Feautrier}, {Ferruzzi}, {Forveille}, {Fossat}, {Foy}, {Fraix-Burnet},
  {Gallardo}, {Giani}, {Gil}, {Glentzlin}, {Heiden}, {Heininger}, {Hernandez
  Utrera}, {Kamm}, {Kiekebusch}, {Le Contel}, {Le Contel}, {Lesourd}, {Lopez},
  {Lopez}, {Magnard}, {Marconi}, {Mars}, {Martinot-Lagarde}, {Mathias},
  {M{\`e}ge}, {Monin}, {Mouillet}, {Mourard}, {Nussbaum}, {Ohnaka}, {Pacheco},
  {Perrier}, {Rabbia}, {Rebattu}, {Reynaud}, {Richichi}, {Robini},
  {Sacchettini}, {Sch{\"o}ller}, {Solscheid}, {Spang}, {Stee}, {Stefanini},
  {Tallon}, {Tallon-Bosc}, {Tasso}, {Testi}, {Vakili}, {von der L{\"u}he},
  {Valtier}, {Vannier}, {Ventura}, {Weis}, \& {Wittkowski}}]{Weigelt2007}
{Weigelt}, G., {Kraus}, S., {Driebe}, T., {et~al.} 2007, \aap, 464, 87

\end{thebibliography}

\end{document}